# Interplanetary radio type II and type IV bursts as indicators of propagating solar transients

Silja Pohjolainen*[1] and Nasrin Talebpour Sheshvan[2]
(1) Tuorla Observatory, University of Turku, Piikkiö, Finland
(2) Space Research Laboratory, University of Turku, Turku, Finland

## Abstract

Recent studies of interplanetary radio type II bursts and their source locations are reviewed. As these bursts are due to propagating shock waves, driven by coronal mass ejections, they can be followed to near-Earth distances and can be used to predict the arrival times of geo-effective disturbances. Radio type IV bursts, on the other hand, are usually due to moving magnetic structures in the low corona and trapped particles form at least part of the emission. The observed directivity of type IV emission may also be used for space weather purposes.

## 1 Introduction

Solar events, mainly flares and coronal mass ejections (CMEs), are associated with efficient particle acceleration that can be observed in radio emission. Accelerated particles either cause oscillations in the surrounding medium, creating radio emission at the local plasma frequency, or trapped particles gyrate in the magnetic field and emit synchrotron radiation. The two different emission mechanisms can be identified, for example, from their flux spectrum and frequency range.

Solar radio bursts are traditionally classified as type I (accelerated particles causing radio storms), type II (propagating shock fronts that accelerate particles), type III (fast-moving particle beams), and type IV (moving magnetic structures that may also contain trapped particles). Types I – III can be identified by their frequency-drifting emission, as the plasma frequency decreases when the transients are moving outward from the Sun, to lower density plasma. Type IV bursts are more complicated to interpret, as they can contain both plasma and syncrotron emission.

Not all solar events create radio emission, some are radio-loud and some radio-quiet, and there are differences in their geo-effectiveness, see for example [1] and the references therein. These differences can be associated with shock speeds, propagation directions, transient interaction that may change CME/shock speeds during propagation, and other relevant features that can be used for space weather purposes.

Interplanetary radio bursts at decameter–hectometer (DH) waves can be observed from space and during 2007–2014 there were three instruments recording solar radio emission from three different spacecraft, Wind, STEREO A, and STEREO B. Figure 1 shows an example of a solar event observed on 22 September 2011, with the three different spacecraft providing simultaneous radio dynamic spectra and a whole 360-degree view around the Sun.

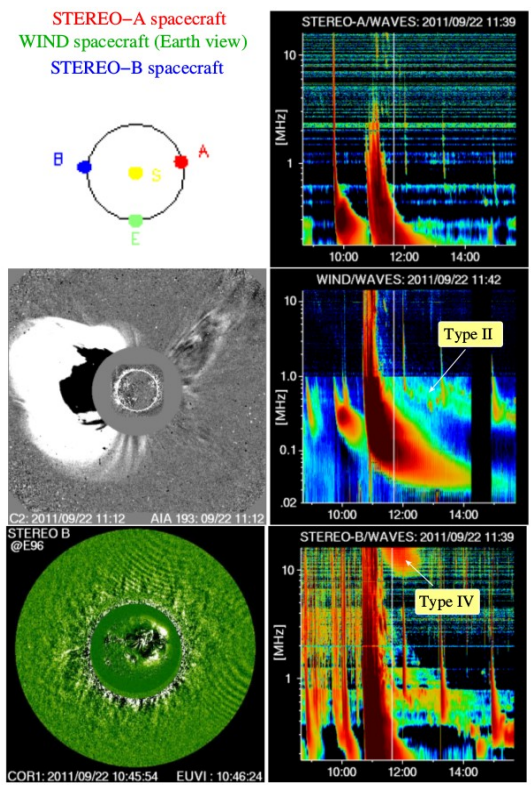

**Figure 1.** Solar radio emission observed with STEREO A and B/SWAVES and Wind/WAVES instruments, from three different angles around the Sun. STEREO-B observes the event on the disk and a type IV burst is visible in the radio dynamic spectrum. From Earth view this is a limb event, and only a type II burst is visible in the Wind/WAVES spectrum. Both types of radio bursts are undetected by STEREO-A/SWAVES.

## 2 Shock Propagation and Type II Bursts

Radio type II bursts can be imaged at decimetric–metric wavelengths with ground-based radioheliographs and most recently with interferometers like LOFAR [2], but for the longer wavelengths that appear in the interplanetary (IP) space indirect methods have to be used to determine source locations. *Direction-finding* technique, see e.g. [3], can be applied when observations are done with one single spacecraft. When more than one spacecraft are available, with sufficiently different viewing angles, *radio triangulation* is possible. As triangulation can be quite complicated, only a few solar radio events have so far been analysed using this method [4, 5, 6]. In these cases the type II emission has been found to be associated with the CME-driven shock, located either at the CME flanks [4, 5], or near the CME leading front [6].

The data from different viewing angles from instruments orbiting the Sun can also be used in a more simple way, to analyse the existance/non-existance and intensity of certain emission features, in order to determine their location. For example a study of the radio enhancements in the type II burst emissions [7] revealed that the enhancements can have many different origins, of which CME interaction with streamers and earlier transients are the most typical. Some type II bursts that show an exceptionally wide band of emission, with no observed harmonic emission lanes, may form a different category in their source origins [7, 8].

## 3 Directivity of Type IV Bursts

Moving radio type IV bursts are typically launched at decimetric–metric wavelengths, but some of them extend to hectometric wavelengths (frequencies lower than about 20 MHz). These low-frequency bursts are rather rare, for example a recent statistical study listed only 48 bursts detected in the Wind/WAVES data in 1998-2012 [9].

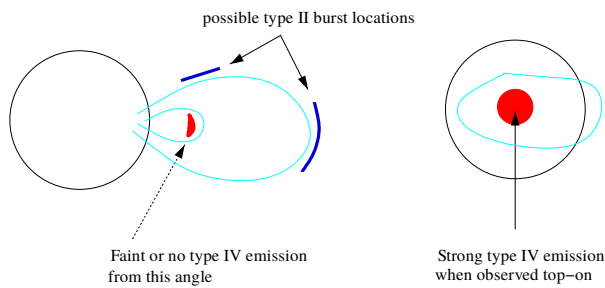

**Figure 2.** Schematic drawing of a possible type IV emission source location, the idea was presented already in [10]. The drawing also shows the typical locations of type II bursts, formed as a bow shock at the leading front of a CME or as a shock at the CME flanks.

Type IV bursts have been known to show directivity effects [10], and simultaneous observations with Wind and STEREO A and B spacecraft have recently confirmed this [11]. A schematic drawing in Figure 2 shows one possible scenario for the observed directivity, to explain the less-intense or missing type IV emission. Strong type IV radio emission is typically observed when a fast, halo-type CME is propagating towards the observer. As the emission mechanism can be both plasma emission and synchrotron emission by mildly relativistic electrons trapped in the magnetic structures, the source origin is not clear. An often-suggested interpretation, at least for metric type IV bursts, is emission from trapped particles in post-eruption loops, after a large flare.

In our ongoing study (Talebpour Sheshvan et al.) we have analysed a selected sample of IP type IV bursts and compared them with the characteristics of the associated flares, coronal waves, and CMEs. A typical interplanetary type IV burst is observed when a GOES X-class flare appears on the solar disk, is associated with an EUV wave, and is followed by a fast, halo-type CME. There is also a strong association with interplanetary type II bursts, although a type IV burst can appear without a type II burst.

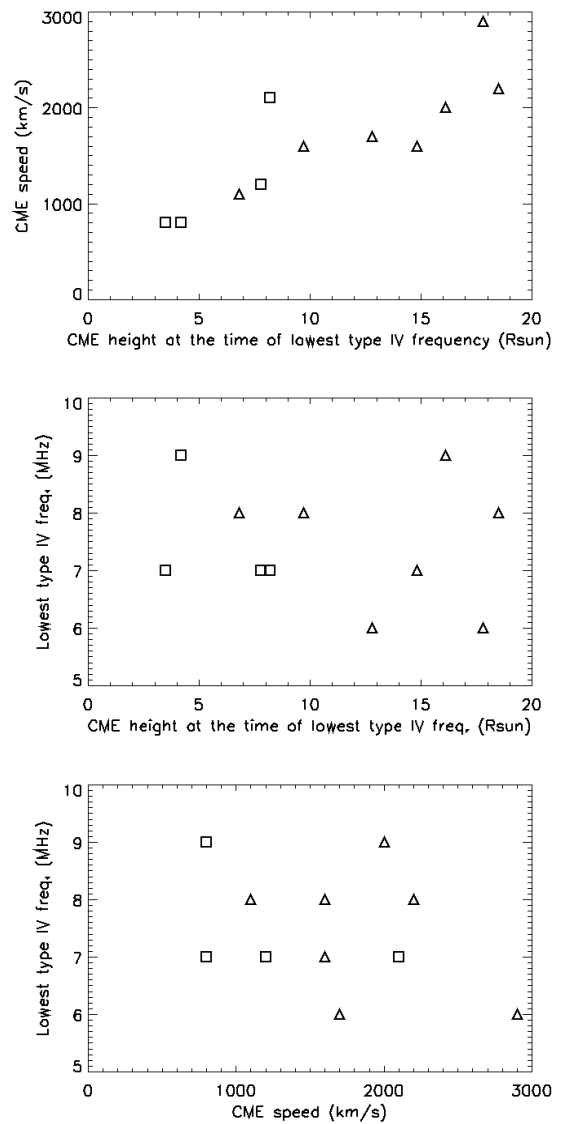

**Figure 3.** Top: CME height at the time of the lowest frequency of type IV emission vs. CME speed. Middle: CME height at the time of the lowest frequency of type IV emission vs. lowest frequency of type IV emission. Bottom: CME speed vs. lowest frequency of type IV emission. Triangles indicate full and intense type IV bursts and squares faint and partial type IV bursts.

Correlation plots in Figure 3 show how the type IV frequency range is almost independent of the CME speed. However, in order to create type IV emission at DH wavelengths the CME must have relatively high speed. The CME height at the time when the type IV burst reaches it's lowest frequency also varies. The lowest observed type IV emission frequencies were 6 – 9 MHz, which correspond to heliocentric heights of about 2 – 4 $R_{sun}$ (the height range also depends on the selected atmospheric density model). This means that there is typically a large spatial separation between the CME leading front and the type IV burst source.

All the CMEs and type IV bursts were associated with EUV waves. The EUV wave speeds were in the range of 542 – 1016 km s$^{-1}$ (speeds obtained from [12]), and no clear correlation was found between the EUV wave speeds and the type IV frequencies.

## 4 Conclusions

Solar transients create radio signatures during their propagation in the interplanetary medium, and identifying these can help to predict shock arrivals and geomagnetic storms. During periods of high solar activity several CMEs can be launched close to each other, both spatially and temporally, and therefore interactions can occur and the effects can sometimes be observed directly in the radio emission. The 3D-view provided by the radio instruments onboard STEREO A, STEREO B, and Wind spacecrafts has helped us to understand how the features observed at DH wavelengths are associated with the propagating disturbances in the interplanetary space.

## 5 Acknowledgements

The radio dynamic spectra, EUV images, and CME data can be obtained from the CME catalog that is generated and maintained at the CDAW Data Center by NASA and The Catholic University of America in cooperation with the Naval Research Laboratory. N. Talebpour Sheshvan wishes to thank CIMO, Finland, for their financial support.